\documentclass[12pt]{article}
 
\def\be{\begin{equation}}
\def\ee{\end{equation}}
\def\a{\alpha}
\def\e{\epsilon}
\def\s{\sigma}

\def\G{\Gamma}
\def\D{\Delta}
\def\d{\Delta}
\def\Tr{\mbox{Tr}}
\def\vac{|0\rangle} 
\def\ra{\rangle}
\def\la{\langle}
\def\Re{\mbox{Re}}
\def\diag{\mbox{diag}}
\def\sin{\mbox{sin}}
\def\sh{\mbox{sh}}
\def\ch{\mbox{ch}}
\def\th{\mbox{th}}

\def\ln{\mbox{ln}}
                       
\begin{document}

\begin{center} 
{\large On the exactly-solvable pairing models for bosons.}
\end{center}

\begin{center}
{\large  An.A. Ovchinnikov  }
\end{center}

\begin{center}
{\it Institute for Nuclear Research, RAS, Moscow, 117312, Russia} 
\end{center}

\vspace{0.1in}

\begin{abstract}

We propose the new exactly solvable pairing model for bosons 
corresponding to the attractive pairing interaction. 
Using the electrostatic analogy, the solution of this model in 
thermodynamic limit is found. 
The transition from the superfluid phase with the Bose condensate 
and the Bogoliubov - type spectrum of excitations  
in the weak coupling regime to the incompressible phase with the gap 
in the excitation spectrum in the strong coupling regime is observed.  

\end{abstract}

\vspace{0.1in} 

              {\bf 1. Introduction.}  

\vspace{0.1in} 

At present time the discrete-state BCS-type \cite{BCS} pairing models 
attract much attention mainly in connection with the physics of 
ultra-small metallic grains (for a review see \cite{BD}). 
Previously, the discrete-state BCS model was solved by Richardson \cite{RS} 
in the context of nuclear physics. Later the integrability of the model was 
shown in ref.\cite{CRS}.   
      The BCS- type exactly solvable discrete pairing model for the system 
of bosons was first considered by Richardson \cite{R68}. 
In the continuum limit the condensate fraction and the Bogoliubov type 
spectrum of the low energy excitations (phonons) was obtained. 
Recently in ref.\cite{DS}, \cite{DES} the pairing model for bosons in the 
context of finite system of bosons confined to a trap was considered. 
For this problem several generalizations of the simplest pairing model 
analogous to the BCS (fermionic) case (for example, see \cite{ADO})
have been proposed. The remarkable feature of the pairing models for the 
confined bosons \cite{DS} is the phenomenon of Bose - condensation.    
In both cases at certain conditions another interesting 
phenomenon of the condensate fragmentation have been observed. 
      From the theoretical point of view, the BCS-type pairing models  
are of interest due to their connection with the generalized Gaudin 
magnets, Knizhnik-Zamolodchikov equations, conformal field theory, and 
Quantum Inverse Scattering method (for example, see \cite{FST}) 
which was studied in a number of papers \cite{G}, \cite{Sierra}. 
From the mathematical point of view the models \cite{R68}, \cite{DS}  
corresponds to the non-compact group $SU(1,1)$ which is the particular case 
of general non-compact $SU(n,m)$ groups \cite{Barut}.  
The construction is equivalent to the 
non-compact $SL(2,R)$ spin chain with different infinite- dimensional 
representations at different sites.

In the present paper we propose and solve the simple modification  
of the model \cite{R68} corresponding to the attractive pairing 
interaction. Naively, for the attractive pairing interaction the ground 
state energy is a decreasing function of the particle number.  
However, one can consider the simple modification of the model \cite{R68} 
which has the correct behaviour of the ground state energy both for the 
finite system and in the continuum limit. 
Namely, for the infinite - volume system of bosons  
we consider the Hamiltonian 
\be
H=\sum_{p}\e_{p}n_{p}-g\sum_{p,p'}(a_p^{+}a_{-p}^{+})(a_{p'}a_{-p'})  
+g'\sum_{p,p'}(a_p^{+}a_{p})(a_{p'}^{+}a_{p'}),    
\label{inf}
\ee
where $g>0$, $n_p=a_p^{+}a_p$, $a_p^{+}$ ($a_p$) - are Bose creation 
(annihilation) operators corresponding to the plane waves with the momentum 
$p_a=2\pi n_a/l$ ($a=x,y,z$, $l$- is the linear size of the system) and 
$\e_p$ is the dispersion for the free particles  
(for example, $\e_p=p^2/2m$).  For pairing models for bosons 
confined  in the external potential the indices in the Hamiltonian (\ref{inf}) 
should represent the states  with definite principal quantum number and  
angular momenta \cite{RS}.  
For the system in the thermodynamic limit the sum over momenta $p$, $p'$ 
in the second term of eq.(\ref{inf}) should be restricted to the values 
$|p|,|p'|<P$, where $P$ is some cutoff $P\sim V$, and in order to have the 
correct behaviour of the ground state energy in the continuum limit, 
the constant $g$ should be rescaled as $g\to g/V$, $V=l^3$.  
Although the last term in eq.(\ref{inf}) is nothing else but 
the constant equal to $g'N_b^2$, where $N_b$ is the 
fixed number of bosons, the model have 
the correct ground state and in many aspects is a more realistic 
one in comparison with the model with repulsion considered by 
Richardson \cite{R68}.  
The model (\ref{inf}) can be applied both for the finite system 
of confined bosons and for the system in the thermodynamic limit. 
Note, that if one considers the model (\ref{inf}),  
as a result of truncation of the initial realistic interaction,  
in general, the terms of both type should be included. 
Note also that in the real systems like $\mbox{He}^4$ the 
attractive tail of the potential at large distances 
is always exist. For the finite systems, if the pairing interaction 
is considered as a residual interaction, the coupling constant 
can be of either sign.    
Previously, the modification of the model (\ref{inf}) 
for the case of attraction was studied numerically for the particular 
values of the parameters by Dukelsky and Schuck \cite{DS}.

For the model (\ref{inf}) in the limit of the large number of bosons 
we find the excitation spectrum and the occupation probabilities 
for an arbitrary value of the coupling constant. 
As a function of the coupling constant $g$  
we observe the discontinuous transition between 
the two different regimes for the model (\ref{inf}). 
In the weak coupling regime there is the Bose condensate and the 
Bogoliubov-type spectrum of excitations. 
In the strong coupling limit the condensate is absent 
and there is a gap in the excitation spectrum.  
Qualitatively the results does not depend on the spacing 
and degeneracy of the energy levels and are valid in the limit 
of the large number of particles. 
The results are compared with the predictions of the 
mean- field theory approach and the Bogoliubov approximation. 
For the attractive model the naive mean - field approximation 
gives the exact results in thermodynamic limit in the case 
of the strong coupling, while the Bogoliubov approximation is exact 
in the weak coupling limit in some range of density depending 
on the coupling constant $g$.    
  We show that the mean-field (variational) approach 
can be modified in order to take into account the 
Bose condensate and can be used for the model (\ref{inf}) 
to obtain the exact results in the thermodynamic 
limit in the whole range of parameters.

In Section 2 we review the exact solution of the model, and 
present the known generalizations of the model.  
We show that this class of models for bosons naturally 
appears in the quasiclassical limit of the algebraic Bethe ansatz 
transfer matrix. We also present some new generalizations 
of the model \cite{R68} which can be useful for studying the 
superfluidity in the framework of this model.  
In Section 3 we present the solution of the model (\ref{inf}) 
in thermodynamic limit.  
In Section 4 we compare the exact solution with the predictions 
of the mean-field theory (variational) approach.

\vspace{0.2in}

     {\bf 2. Pairing models for bosons.}

\vspace{0.1in}

The Hamiltonian for the boson pairing model has the form 
\be
H=\sum_{i=0}^{L-1}\e_i n_i+ gB^{+}B^{-}, 
\label{H}
\ee
where the coupling constant $g$ is positive for the repulsion. 
The operators in (\ref{H}) are defined through 
the pairs of bosonic creation and annihilation operators 
$\phi_{1i}^{+}$, $\phi_{2i}^{+}$ at the $i$th energy level $\e_i$ 
($\left[\phi_{\s i};\phi_{\s i}^{+}\right]=1$, $\s=1,2$).   
In terms of the pair creation and annihilation operators 
\[
b_i^{+}=\phi_{1i}^{+}\phi_{2i}^{+},~~~~~~~b_i=\phi_{1i}\phi_{2i},  
\]
the operators $B^{\pm}$ are defined as $B^{+}=\sum_{i}b_i^{+}$, 
$B^{-}=\sum_{i}b_i$, and $n_i=n_{1i}+n_{2i}$ 
($n_{\s i}=\phi_{\s i}^{+}\phi_{\s i}$). 
For each energy level $\e_i$ the operators $\phi_{\s i}^{+}$ 
describe the pair of states differing by the time reversal symmetry.  
For example, for the translationally invariant 
system the pair creation operator has the form  
$b_p^{+}=\phi_{p}^{+}\phi_{-p}^{+}$ (zero total momentum). 
An arbitrary degeneracy $\Omega_i$ of each energy level can also 
be taken into account. 
   For the correct behaviour of the ground state one should re-scale 
$g\to g/L$, where $L$ is the number of sites, in eq.(\ref{H}).    
The rescaled value of $g$ will be substituted in the final results 
throughout the paper.  
We define the particle density $\rho=N_b/L$, where $N_b$ is the 
total number of bosons. 
   The commutational relations for the pair creation and annihilation 
operators have the form
\[
\left[(\phi_{1i}\phi_{2i});(\phi_{1i}^{+}\phi_{2i}^{+})\right]=1+n_i,~~~~~
n_i=\phi_{1i}^{+}\phi_{1i}+\phi_{2i}^{+}\phi_{2i}. 
\]
The commutational relations with the operator of the number of bosons 
$n_i$ are: 
\[
\left[b_i^{\pm};n_i\right]=\mp 2 b_i^{\pm}.  
\]
These commutational relations are equivalent to the group algebra 
of the pseudo-spin generators for the group $SU(1,1)$, which differs 
by the sign for the commutator $\left[ S_i^{+};S_i^{-}\right]$ 
from that of the $SU(2)$ algebra \cite{Barut}. 
As in the case of the conventional discrete BCS-type models, 
the eigenstates can be constructed directly in terms of the operators 
\[
\Sigma^{+}(t)=\sum_{i}\frac{b_i^{+}}{(t-\e_i)},~~~~~~
b_i^{+}=\phi_{1i}^{+}\phi_{2i}^{+}.  
\]
We seek for the eigenstates of the Hamiltonian (\ref{H}) in the form:   
\be
|\phi(t)\ra=\Sigma^{+}(t_1)\Sigma^{+}(t_2)\ldots\Sigma^{+}(t_M)|\nu\ra,
\label{t}
\ee
where the state $|\nu\ra$ contains only the unpaired states i.e. defined 
by the conditions: 
\[
b_{i}|\nu\ra=(\phi_{1i}\phi_{2i})|\nu\ra=0, 
~~~~~~   n_{i}|\nu\ra=\nu_{i}|\nu\ra,  
\]
where $\nu_i$ are the (conserved) numbers of the unpaired bosons at the 
level $i$. Note that the state (\ref{t}) is degenerate and does not 
determine the eigenstate completely. One should introduce the 
additional quantum numbers $\sigma_i=\pm1$ such that for each site 
$(n_{1i}-n_{2i})|\nu\ra=\sigma_i\nu_{i}|\nu\ra$. 
The energy does not depend on $\sigma_i$, but the (angular) momentum 
depends on the quantum numbers $\sigma_i$. 
The complete set of states is given by the formula 
\[
\left( (b_0^{+})^{n_0}(b_1^{+})^{n_1}\ldots(b_{L-1}^{+})^{n_{L-1}}\right) 
|(\nu_0,\sigma_0),(\nu_1,\sigma_1),\ldots 
(\nu_{L-1},\sigma_{L-1})\ra, 
\]
and can be characterized by integer quantum numbers $n_i$, $\sum_{i}n_i=M$, 
instead of the parameters $t_i$ (apart from $\nu_i$, $\sigma_i$). 
Since at $g\to0$ the Hamiltonian reduces to the free one each eigenstate 
of (\ref{H}) at finite $g$ can be characterized by the integers 
$n_i^{(0)}$, $\sum_{i}n_i^{(0)}=M$, corresponding to the state at $g=0$. 
In the limit $g\to0$ the set of $n_i^{(0)}$ parameters $t_j\to\e_i$.  
We use the following commutational relations for the operator 
$\Sigma^{+}(t)$ which can be proved using the commutational 
relations for $b_i^{+}$, $b_i$, $n_i$, and different from the  
formulas for the spin- $1/2$ case:  
\be
n_i\Sigma^{+}(t)=\Sigma^{+}(t)n_i+\frac{2}{(t-\e_i)}b_i^{+}, ~~~~~ 
b_i\Sigma^{+}(t)=\Sigma^{+}(t)b_i+\frac{1}{(t-\e_i)}(1+n_i). 
\label{com} 
\ee
For the Hamiltonian (\ref{H}) the equations for the parameters $t_i$ 
(\ref{t}) are obtained in the same way as the formulas for the 
generalized Gaudin magnets (for example, see \cite{R68}, \cite{G}). 
The energy eigenvalues and the equations for the eigenstates are: 
\be
E=\sum_{i}\e_{i}\nu_i+2\sum_{i=1}^{M}t_i, ~~~~~~~~
\sum_{\a}\frac{\Omega_{\a}+\nu_{\a}}{t_i-\e_{\a}}
+\sum_{j\neq i}\frac{2}{t_i-t_j}=\frac{2}{g}. 
\label{eq}
\ee 
The total number of bosons equals: $N_b=\sum_{i=0}^{L-1}\nu_i+2M$. 
Note that for the Hamiltonian (\ref{H}) the number of pairs and 
the number of the unpaired particles is conserved.  
Since the operator $\Delta n_i=n_{1i}-n_{2i}$ commutes with 
the generators of $SU(1,1)$ group one can add the term 
$\sum_{i}h_{i}\Delta n_i$ to the Hamiltonian (\ref{H}) 
to obtain the model with the external field $h_i$. 
In this case the equations for $t_i$ are the same 
as for the model (\ref{H}), while the energy equals  
$E=\sum_{i}(\e_i+h_i\sigma_i)\nu_i+2\sum_{i}t_i$.  
The equations (\ref{eq}) are different from the equations for the BCS  
case by the normalization factors and the sign of the second term 
at the left-hand side.  
In the same way  as in ref.\cite{CRS}, the set of commuting operators 
$H_i$ ($i=1,\ldots L$) and their eigenvalues can be found. 
In fact, analogously to the case of the $SU(2)$ group, consider 
the operators: 
\be
H_i=\frac{1}{g}n_{i}+\sum_{l\neq i}\frac{(S_{i}S_{l})}{(\xi_i-\xi_l)}
\label{Hi}  
\ee
where we have denoted by $(S_{i}S_{j})=\sum_{a=x,y,z}S_{i}^{a}S_{j}^{a}$ and 
defined 
\be
S_{i}^{+}=ib_i^{+},  ~~~ S_{i}^{-}=ib_i, ~~~ S^z=\frac{1}{2}(1+n_i).  
\label{spins}
\ee
Note that in terms of initial 
$SU(1,1)$ generators $b_i^{+}$, $b_i$, $1+n_{1i}+n_{2i}$, 
the scalar product has the form: 
\[
(S_{i}S_{j})=-\frac{1}{2}(b_i^{+}b_j+b_j^{+}b_i)+\frac{1}{4}(1+n_i)(1+n_j). 
\]
Since the commutational relations for the operators $S_{i}^{a}$  
are the same as for the $SU(2)$ group, in analogy with the discrete - 
state BCS- model \cite{CRS}, \cite{DS}, the operators (\ref{Hi}) commute 
$\left[H_i;H_j\right]=0$. 
At $\e_i=\xi_i$ the linear combination $\sum_{i}\e_{i}H_i$ gives the 
Hamiltonian (\ref{H}) while for general $\e_i\neq\xi_i$ we obtain the 
Hamiltonian depending on the two sets of parameters: 
\be
H=\sum_{i}\e_{i}n_{i}+ 
g \sum_{i<j}\frac{(\e_i-\e_j)}{(\xi_i-\xi_j)}(S_{i}S_{j}).
\label{exi}
\ee
It was noted in ref.\cite{DS} that the choice $\xi_i=(\e_i)^{d}$ leads to the 
realistic model for bosons confined in $d$ - dimensional magnetic trap 
represented by the external harmonic well potential. 
The equations determining the eigenvalues of the Hamiltonian (\ref{exi}) 
and the operators (\ref{Hi}) are given by the equations (\ref{eq}) 
with the parameters $\e_i$ replaced by $\xi_i$.

Let us comment on the inclusion of the energy level which 
corresponds to the single Bose creation operator $\phi_0^{+}$ ($p=0$ level 
in the system with periodic boundary conditions or the $n=0$ energy level 
in the spherically symmetric system) i.e. of the non-degenerate energy level. 
One can formally consider the states build up with {\it two} Bose creation 
operators of the form $(\phi_1^{+}\phi_2^{+})^{n}|\nu_0\ra$, where 
$\nu_0=0,1$, and associate with this state the state $|\nu_0+2n\ra$ of 
$\nu_0+2n$ bosons at the energy level $0$. The interaction  
with the other pairs remains the same i.e. of the type  
$(\phi_1^{+}\phi_2^{+})(\phi_{1i}\phi_{2i})$ ($i\neq0$). Thus the energy level 
$n=0$ is considered on equal footing with the other energy levels with the 
exception of the allowed value of the parameter $\nu_0=0,1$   
which corresponds to the special type of interaction of pairs  
with the particles at the energy level $0$.

Let us show that the discrete - state bosonic pairing models 
presented above as well as the new models with the interaction of pairs 
depending on the energy levels, can be obtained in the framework of the 
Quantum Inverse Scattering Method (for example, see \cite{FST}). 
Consider the Monodromy matrix defined in the usual way: 
\[
T_0(t)=K_{0}L_{10}L_{20}\ldots L_{N0}, 
\]
where $K_{0}=\diag(e^{-\eta/2g};e^{\eta/2g})$ is the usual diagonal twist 
matrix and the Lax operator obeying the Yang-Baxter equation is given by 
\be
L_{i0}(\xi_i,t)=\xi_i-t+\eta(\s S_i), 
\label{Lax}
\ee
where the operators $S^a_i$ ($a=x,y,z$) were defined through the $SU(1,1)$ 
generators in the previous section, 
$\s^a$ are the Pauli sigma matrices and 
$\xi_i$ are the inhomogeneity parameters. 
Considering the quasiclassical limit $\eta\to0$ 
of the transfer matrix $Z(t)=\Tr_{0}(T_0(t))$, we obtain the family of the 
Hamiltonians depending on the spectral parameter $t$, 
which commute at different values of the parameters:   
\be
H(t)=\frac{1}{2g}\sum_{i}\frac{1}{(t-\xi_i)}(1+n_i)
+2\sum_{i<j}\frac{1}{(t-\xi_i)(t-\xi_j)}(S_{i}S_{j}),    
\label{Ht}
\ee  
$\left[H(t);H(t')\right]=0$, where the expression for the scalar 
product $(S_{i}S_{j})$ was presented in the previous section. 
From eq.(\ref{Ht}) taking the limits $t\to\xi_i$ or 
$t\to\infty$ the different pairing models for bosons can be obtained. 
The limit $t\to\xi_i$ corresponds to the operator $H_i$ (\ref{Hi}). 
The eigenvalues of $H(t)$ can be obtained from the eigenvalues 
of the transfer matrix $Z(t)$.  
We define the monodromy matrix in the following way: 
\[
T_0(t)=\left(
\begin{array}{cc}
A(t) & B(t) \\
C(t) & D(t) 
\end{array} \right)_0,  
\]
and seek for an eigenstates in the form 
\[
|\phi(t)\ra= B(t_1)B(t_2)\ldots B(t_M)|\nu\ra, 
\]
where the reference state $|\nu\ra$ is defined by the conditions 
$S^{-}_i|\nu\ra=0$ and $n_i|\nu\ra=\nu_i|\nu\ra$. 
As in the usual $SU(2)$ case we observe that 
\[
C(t)|\nu\ra=0, 
\]
and the state $|\nu\ra$ is an eigenvector of $A(t)$ and $D(t)$.  
The eigenvalues of the operators $A(t)$ and $D(t)$ are: 
\[
A(t)|\nu\ra=\prod_{\a}\left(\xi_{\a}-t+\eta(1+\nu_{\a})/2\right)|\nu\ra, ~~~
D(t)|\nu\ra=\prod_{\a}\left(\xi_{\a}-t-\eta(1+\nu_{\a})/2\right)|\nu\ra. 
\]
Following the well known procedure we obtain the Bethe ansatz equations: 
\be
e^{\eta/g}\prod_{\a=1}^N\left(
\frac{t_i-\xi_{\a}+\eta(1+\nu_{\a})/2}{t_i-\xi_{\a}-\eta(1+\nu_{\a})/2}\right)
= \prod_{\a\neq i}^M\left(\frac{t_i-t_{\a}-\eta}{t_i-t_{\a}+\eta}\right)
\label{equa} 
\ee
The corresponding eigenvalue of the transfer matrix $Z(t)$ equals  
\[
\Lambda(t)= 
\prod_{i}\left(\frac{t_i-t+\eta}{t_i-t}\right)
\prod_{\a}(\xi_{\a}-t+\eta(1+\nu_{\a})/2)+ 
\prod_{i}\left(\frac{t-t_i+\eta}{t-t_i}\right) 
\prod_{\a}(\xi_{\a}-t-\eta(1+\nu_{\a})/2), 
\]
where the two terms corresponds to the operators $A(t)$ and $D(t)$ 
respectively. Considering the quasiclassical limit of the Bethe ansatz 
equations, one reproduce the equations (\ref{eq}) for the 
eigenstates of the pairing Hamiltonian (\ref{H}). 
The eigenvalues of the operators $H_i$ and $H$ (\ref{H}) 
can be obtained from the last expression for $\Lambda(t)$. 
According to the general procedure \cite{Fadd} 
one can obtain the fundamental $R$ - matrix corresponding to the direct 
product of two representations of the group $SU(1,1)$ 
(the so called fundamental Lax operator)    
which will lead to the new transfer matrix $Z^{(f)}(t)$ with the trace 
over the infinite- dimensional space, commuting with the 
transfer matrix $Z(t)$ and the Hamiltonians of the new type, 
which is beyond the scope of the present paper.  
In order to obtain the trigonometric transfer matrix, 
one should have the special quantum group commutational relations 
$\left[S^{+};S^{-}\right]=\sin(2\eta S^z)/\sin(\eta)$, 
which are not fulfilled for the $SU(1,1)$ generators. 
However, since the commutational relations for $S^{\pm}$, $S^{z}$ 
coincide with the usual $SU(2)$ algebra,  
the Gaudin - type Hamiltonians \cite{ADO},
which are quadratic in the operators $S^a$, 
can be obtained in the trigonometric case. 
Thus all the known results, for the Gaudin- type Hamiltonians for the  
trigonometric and the elliptic cases, can be generalized to the case 
of $SU(1,1)$ - generators, constructed with the help of 
bosonic operators.

Let us briefly comment on the possible generalizations of these models. 
One can use the representation of spin-$s$ operators through the 
single Bose creation and annihilation operators $\phi^{+}$, $\phi$, 
$\left[\phi;\phi^{+}\right]=1$ as $S^{+}=\phi^{+}(2s-\phi^{+}\phi)^{1/2}$, 
$S^{-}=(2s-\phi^{+}\phi)^{1/2}\phi$, $S^z=\phi^{+}\phi-s$,   
to include this spin in the quasiclassical Hamiltonian.  
In the fermionic case this leads to the generalized Dicke model 
if the limit $s\to\infty$ is taken. If one considers 
the limit $\xi_1\to\infty$ for this single site in the hyperbolic 
version of the model (\ref{exi}) one can eliminate 
the terms, which do not conserve the number of bosons and   
obtain the model describing the interaction of 
single oscillator with the bosonic degrees of freedom: 
\[
H=\omega\phi^{+}\phi+
\lambda(\phi^{+}\phi)
\left(\sum_{i}\e_{i}n_i+\sum_{i}c_{i}\s_{i}\nu_i\right)+ 
\sum_{i}h_{i}\nu_i+\sum_{i}\e_{i}n_i+gB^{+}B^{-}.  
\]
This Hamiltonian contains a number of free parameters  
which can be chosen in order to get the realistic model. 
In the sector with the oscillator excited to $n$-th 
energy level the model is reduced to the boson pairing 
model with the renormalized coupling constant $g$. 
At the same time the excitation energy (level spacing) for the 
oscillator depends on the average occupation numbers $n_i$ for  
bosons. In contrast to the same model for fermions, the 
occupation probabilities $\la n_i\ra$ can be a small numbers, 
which allows for the small renormalization of the oscillator 
frequency. This model can be useful for studying the 
superfluidity without any assumptions.

\vspace{0.2in}

  {\bf 3. Continuum limit.}

\vspace{0.1in}

Let us solve the model (\ref{H}) in the continuum limit.     
Assuming that the distribution of roots $t_i$ can be approximated 
by the continuous density $R(t)$, which is valid for the large number 
of pairs $M$, we get the following equation: 
\be
 \int_a^b dt'\frac{R(t')}{t-t'}=f(t), ~~~~
f(t)=\frac{1}{g}-\frac{1}{2}\sum_{\a}\frac{C_{\a}}{t-\e_{\a}}, 
\label{coneq}
\ee
where the integral in a sense of principal value over the support of 
the function $R(t)$ is implied and $C_{\a}=\Omega_{\a}+\nu_{\a}$.  
According to \cite{R68} for the case of repulsion the ground state 
corresponds to the roots $t_i$ located at the interval $(\e_0,\e_1)$. 
One can argue that since at $g\to0$ the ground state 
corresponds to all $t_i\to\e_0$ and the roots $t_i$ cannot 
cross the values $\e_{\a}$ for varying $g$, all $t_i\in(\e_0,\e_1)$. 
Thus one should solve the equation (\ref{coneq}) assuming that 
the support of the function $R(t)$ is the interval $(\e_0,\e_1)$.  
The structure of the ground state for the repulsion and the  
general behaviour \cite{R68} of solutions of the equations (\ref{eq}) 
can be easily seen from the electrostatic analogy. 
   Electrostatic analogy for the equations of the type (\ref{eq}) 
was previously used for the solution of the equations for the 
case of the BCS problem (for example, see \cite{Gaudin}).     
One can consider the functional of the roots $t_i$ 
as an energy of charges at the two-dimensional complex plane:
\[
\Phi(t_i)=-\sum_{i,\a}C_{\a}\ln|t_i-\e_{\a}|
-2\sum_{i<j}\ln|t_i-t_j|+(2/g)\sum_{i}\Re(t_i).  
\]
This energy functional corresponds to the repulsion of unit charges 
$t_i$ and the repulsion of the charges $t_i$ with the charges of the 
same sign $C_{\a}>0$ placed at the fixed points $\e_{\a}$. 
The external electric field of the magnitude $1/g$ is applied.  
The condition of stationary point (not minimum) of the 
functional $\Phi(t_i)$ with respect to the positions of the charges 
$t_i$ leads to the equations (\ref{eq}).  
It is convenient to imagine the charges $\e_{\a}$ placed at the 
$y$ axis of $(x,y)$ plane as an energy levels.  
Then for the case of repulsion the external electric field is directed 
down, and each root gives the contribution to the energy equal to its height. 
One can see that all roots $t_i$ are real, 
since due to the repulsion and the external electric 
field all the other configurations are unstable.  
Physically the picture is as follows. 
For each charge the repulsion due to the external charges $\e_{\a}$ 
below this charge, and the other roots below this 
charge, produce the force directed up. 
This force is compensated by the other charges above this charge 
and the external electric field directed down.  
For the ground state the roots $t_i$ should be placed as low as 
possible.  
This picture allows one to use the physical intuition to find the 
solutions for the ground and the excited states of the model 
(\ref{H}). For instance, the general behaviour of the solutions 
\cite{R68} is obvious.

Here we consider the pairing model (\ref{H}) for the case of attraction 
$g<0$ or equivalently the model (\ref{inf}) for $g>0$.  
It was already mentioned that due to the additional term (\ref{inf}) the 
behaviour of the ground state energy as a function of particle number 
is correct.  
In many aspects the model (\ref{inf}) is more realistic in 
comparison with the model with repulsion \cite{R68}.   
For example, it has the Bogoliubov-type 
spectrum of excitations and the Bose condensate which varies 
continuously with the coupling constant from    
zero at some critical coupling $g_c$ to $N_b$ at $g=0$. 
Later on we will omit the last term in eq.(\ref{inf}) 
in all the formulas.   
   In the framework of electrostatic analogy the case  
of attraction corresponds to the external electric 
field directed up. 
Thus for any coupling constant $g$ for the ground 
state all roots of the equations (\ref{eq}) 
located below the lowest energy level $\e_0=0$. 
For small $|g|$ they are close to $\e_0$, while for large $|g|$ 
they are far below the level $\e_0$. 
At the sufficiently small $|g|$ 
the density of roots $R(t)$ grows at $t\to0$ and bounded 
from below at some fixed point $-b$ ($b>0$).

The simple method to find the solution for $R(t)$ (\ref{coneq})   
is, using the electrostatic analogy, to consider the electric field 
at the complex plane $z$ produced by the unit charges $t_i$ located 
at the interval $\G=(a,b)$ of the real axis, the charges $\e_{\a}$, 
and the external electric field:  
\[
h(z)=\int_a^b dt\frac{R(t)}{z-t}-f(z)  
\]
where the discontinuity $\d h(t)$ at $\G$ is given by 
the density of charges $R(t)$: 
$\d h(t)=h(t+i0)-h(t-i0)=2\pi iR(t)$.   
The equation (\ref{coneq}) takes the form $\bar{h}(t)=0$, where 
$\bar{h}(t)$ is an average value of the field at both sides of $\G$,    
and can be represented in the form: 
\be
\frac{1}{2\pi i}\oint_{C}dz\frac{h(z)}{z-t}=f(t)  
\label{cont}
\ee
for $t\in\G$, where the contour $C$ encloses the interval $(a,b)$. 
For the sufficiently small coupling constant 
we use the following ansatz for the electric field $h(z)$ in the 
complex plane which has the branch cut along the interval $(a,b)$  
(in this case we take $a=0$ and the interval $(-b,0)$, $b>0$ and use 
the coupling constant for the attraction $g>0$):  
\be
h(z)=\sqrt{\frac{z+b}{z}}\left(\int_{-b}^{0}
d\xi\frac{\phi(\xi)}{z-\xi}
-\frac{1}{g}\right), 
\label{any} 
\ee
where the function $\phi(\xi)$ can be fixed from the condition 
for the residues of $h(z)$ at the points $\e_{\a}$ which are equal 
to $-C_{\a}/2$, 
\[
\phi(\xi)=\left(\frac{\xi}{\xi+b}\right)^{1/2}\rho(\xi), ~~~~ 
\rho(\xi)=-\frac{1}{2}\sum_{\a}C_{\a}\delta(\xi-\e_{\a}).
\]
The constant term in the parenthesis is fixed from the behaviour of the 
field at infinity, and the value of $b$ is determined from  
the condition $\int dtR(t)=M$. 
The number of pairs and the 
energy $\D E=\sum_{i}2t_i$ can be represented as an integrals 
in the complex plane over the contour $C$ enclosing the interval $\G$: 
\be
M=-\oint_{C}\frac{dz}{2\pi i}h(z), ~~~~~ 
\D E=-\oint_{C}\frac{dz}{2\pi i}2zh(z). 
\label{me}
\ee
The integrals can be evaluated by means of deformation of  
the contour $C$ into the small contours around the points $\e_{\a}$  
and the large circle at the infinity. 
The equivalent way to find the energy is to substitute 
the ansatz for $h(z)$ into the equation (\ref{cont}), which after 
the deformation of the contour $C$ allows one to find  
the function $\phi(\xi)$, presented above 
and the term $1/g$ in eq.(\ref{any}). 
Using the same formulas for $M$ and $E$ (\ref{me}),  
we obtain the following equation for the particle number:  
\be
\frac{b}{g}=N_b+L-\sum_{\a}C_{\a}S(\e_{\a}), 
~~~~~~~S(\xi)=\left(\frac{\xi}{\xi+b}\right)^{1/2}, 
\label{m}
\ee
which determines the value of the parameter $b$. 
In contrast to the case of repulsion apriory we did not 
have any condition, which determines the upper bound for $|b|$: 
the support of $R(t)$ is not bounded from below for $g\to\infty$.    
The value of $b$ found from the last equation should be substituted 
to the equation for the energy (\ref{me}) which takes the form: 
\[
E=-\sum_{\a}\e_{\a} 
+\sum_{\a}C_{\a}S(\e_{\a})(\e_{\a}+b/2)
-\frac{b^2}{4g}. 
\] 
Using the equation (\ref{m}) one can represent the last 
equation in a more convenient form: 
\be
E=-\sum_{\a}\e_{\a}+\sum_{\a}C_{\a}E(\e_{\a})
-\frac{b}{2}(N_b+L)-\frac{b^2}{4g}, ~~~~~~
E(\e)=\sqrt{\e(\e+b)}.  
\label{te}
\ee
In order to find the excitation spectrum and the occupation 
probabilities one should calculate the 
variation of the energy (\ref{te}) over the quantum numbers 
$\nu_{\a}$ and the energy levels $\e_{\a}$ respectively, 
taking into account the variation of the parameter $b$  
eq.(\ref{m}). 
Let us note that the units 
for $\e_i$ can be chosen in an arbitrary way, 
see eq.(\ref{eq}). The possible choice is 
$\e_1=1/L$, such that $L\e_1=1$. 
In thermodynamic limit there are two parameters - 
the density $\rho$ and the coupling constant $g$. 
For example, one can imagine a one-dimensional lattice  
model with linear dispersion relation and $L$ lattice sites. 
We will assume the units $L\e_1=1$ 
and, as an example, consider the equal-spacing $L$ level 
model with $\Omega_{\a}=1$ and use the rescaled coupling 
constant $g\to g/L$ in the final results. 
We obtain from the equation (\ref{m}) at $\nu_{\a}=0$ the 
following equation for the parameter $b$: 
\be
b=g(f(b)+\rho), 
\label{bg}
\ee
where $f(b)$ is a smooth function which varies 
from zero to unity for $b\in(0,\infty)$.  
For example, for the equal-spacing model with 
$L\e_1=1$ we have 
\[
f(b)=b\ln\left((1+\sqrt{1+b})/\sqrt{b}\right) 
+1-\sqrt{1+b}. 
\]
Equation (\ref{bg}) gives the value of $b$ as a 
function of the parameters $g$, $\rho$. 
   First, consider the limit of the small coupling 
constant $g\ll1$, $g\rho\ll1$, such that $g\ll g\rho$.     
 According to the last formula this limit corresponds to 
$b=g\rho$, and the high density limit $\rho\sim 1/g$.  
In this limit one can neglect the last sum in eq.(\ref{m}) 
and  obtain the excitation spectrum and the occupation 
numbers. Considering the energy (\ref{te}) as a function of 
the quantum numbers $\nu_{\a}$ and taking into 
account the variation of the parameter $b$ according 
to eq.(\ref{m}), we find the spectrum of phonons:  
\[
E(\e)=\sqrt{\e(\e+g\rho)}.  
\]
As in ref.\cite{R68} one can show that the states  
corresponding to the excitation of pairs have the same 
energy, so that the (bosonic) quasiparticle interpretation 
of the excited states is true. 
This formula, corresponding to the particular limit 
$g\ll g\rho \ll 1$, is in agreement with predictions 
of the Bogoliubov approximation.  
However, in contrast to the repulsion, this spectrum 
is exact for an arbitrary value of the parameter $g\rho$ 
with respect to the energy level spacing $\e_1$, provided 
the condition $g\ll1$ is satisfied.

   Variation of the energy (\ref{evar}) with respect to the 
parameters $\e_{\a}$ gives the occupation probabilities 
$\la n_{\a}\ra$ which are different for 
$\la n_0\ra$ (condensate) and $\la n_i\ra$, 
$i\neq0$, which can be easily seen from the electrostatic 
analogy. In fact, if the parameter $g\rho$ is not too large,  
shifting the level $\e_0=0$ down will obviously shift 
the distribution of roots and the lower boundary $-b$ 
down as a whole, which means the existence of the 
condensate. 
Considering the variation $\delta E/\delta\e_i$ for 
$i\neq0$, we obtain: 
\be
\la n_i\ra=\frac{\e_i+g\rho/2}{\sqrt{\e_i(\e_i+g\rho)}}-1, 
~~~~~~i\neq0. 
\label{ni}
\ee
The condensate fraction $N_0$ can be evaluated using the 
equation $N_0=N_b-\sum_{i\neq0}\la n_i\ra$. 
  At $\e_1\ll g\rho$ the sum in (\ref{ni}) can be replaced 
by the integral, which gives: 
\[
N'=N_b-N_0=L\left(\sqrt{1+g\rho}-1\right),  
\]
The parameter which governs the condensate fraction is $g$: 
in the limit considered, $N'=(g\rho)L=gN_b\ll N_b$.

Next, consider the case $b\sim1$. 
According to eq.(\ref{bg}) it is possible in 
the two cases: ($i$) $g\ll1$ and $\rho\sim 1/g\gg1$;  
($ii$) $g\sim1$, $\rho\sim1$. In both cases  
calculating the excitation spectrum and the 
occupation numbers from the equations 
(\ref{m}), (\ref{te}), i.e. taking the variation of 
the energy (\ref{te}) with respect to $\nu_{\a}$ and 
$\e_{\a}$ (taking into account the variation 
of the parameter $b$ according to eq.(\ref{m})) 
we obtain the expressions  
\[
E(\e)=\sqrt{\e(\e+b)}, ~~~~~
\la n_i\ra=\frac{\e_i+b/2}{\sqrt{\e_i(\e_i+b)}}-1, 
~~~~i\neq0. 
\]
and the expression for the condensate 
\be
N'=N_b-N_0=L(\sqrt{1+b}-1).  
\label{b}
\ee
Since $g\rho\sim1$, in the case of large 
density $\rho\gg1$ (case ($i$)) we will always have 
$N'\ll N_b$. However, in the case ($ii$), $\rho\sim1$, 
for the coupling constant $g$ larger than some 
critical value $g_c$, the last equation gives $N'>N_b$.  
That means that for the sufficiently large coupling 
constant the solution (\ref{any}) is not correct.

Below we will show that at $g>g_c$ the 
solution should be modified. We will also 
show that the critical value $g_c$ is 
determined exactly by the condition $N'(b)=N_b$, 
where $b$ is the solution of the equation (\ref{bg})    
(i.e. we will show that this condition coincides with 
the condition (\ref{gc}), see below). 
   Here let us present the physical arguments, which 
show that at large $g$ the new phase with the gap 
in the excitation spectrum should exist. 
   As a limiting case, consider $L$ (the large number) 
levels glued together. In this case the repulsion 
directed down is strong 
in comparison with the external field directed up  
and there cannot be roots $t_i$ in the vicinity 
of $\e_0=0$. Thus the support of $R(t)$ should be 
located far below $\e_0=0$, at the distance 
of order $\sim gL$. The ground state energy will 
be of order $\sim-gLN_b/2$, and the gap in the 
excitation spectrum will exist.      
This picture is in agreement with the energy 
of the one-level model \cite{R68} with 
$\Omega$ replaced by $L$. 
So, as a first step, one could solve the one-level model 
with $\Omega$ replaced by $L$ and $\nu=0$ at $g\to\infty$, 
which would be the particular case of the general solution. 
In other words, at large $g$ (weak external field) 
the distribution of charges 
will be unstable if the length $|b|\sim g\rho$ 
is much larger than the length $L\e_1$.

Thus, we seek for solution of the equation (\ref{coneq}) 
with the density support at the interval $(b,a)$, $a,b<0$.  
In general one expects that since there is no external charges 
$\e_{\a}$ in the vicinity of the interval $(b,a)$,  
the support of $R(t)$, it should be equal to zero at the 
endpoints. The numerical calculations suggest that for $|a|>0$ 
the function $R(t)$ is, in fact, equal to zero at the points $a$, $b$. 
It might seem that the ansatz for $h(z)$ should be chosen in 
such a way that as limiting case $a=0$ it would contain the  
solution for the interval $(b,0)$ i.e. in the form (\ref{any}) with 
$a\neq0$.  
However, we will show that correct solution reproduce eq.(\ref{any}) 
at $a=0$.  
One can perform the calculations with the field of the type (\ref{any}) 
and find that the parameters 
$a$, $b$ are not completely fixed from the solution 
itself and one finds a number of  
functions $R(t)$ with different energy, which is  
not correct, as can be seen from the electrostatic analogy.  
   Thus let us find the solution of (\ref{coneq}) 
with the density $R(t)$ equal to zero at the 
endpoints of the interval $(b,a)$, $a,b<0$.  
Since the field $h(z)$ should be a constant at the 
infinity, we consider the following function:    
\be
h(z)=\sqrt{(z-b)(z-a)}
\left(\int_{a}^{b}d\xi\frac{\phi(\xi)}{z-\xi}\right),   
\label{bcs}
\ee  
where $\e_0=0$ and $a,b<0$. The points $a$, $b$ should 
be determined from the solution itself. Note that there are 
no poles other than the poles corresponding to the charges 
$\e_i$ in $h(z)$.   
After changing the signs of the parameters 
$a$, $b$, from the equation (\ref{cont}) we find 
\[
\phi(\xi)=-\frac{1}{2}\sum_{\a}S(\xi)\delta(\xi-\e_{\a}),
~~~S^{-1}(\xi)=E(\xi)=\sqrt{(\xi+a)(\xi+b)}   
\]
and simultaneously the condition for the behaviour 
of the field at the infinity: 
\[
\int d\xi\phi(\xi)= 
-\frac{1}{2}\sum_{\a=0}^{L-1}S(\e_{\a})=-\frac{1}{g}, 
\]
or, equivalently, 
\be
\sum_{\a}
\frac{C_{\a}}{\sqrt{(\e_{\a}+a)(\e_{\a}+b)}}=\frac{2}{g}. 
\label{gapbcs}
\ee
The first of the equations (\ref{me}) gives 
\be
\frac{a+b}{g}=N_b+L-\sum_{\a}C_{\a}\e_{\a}S(\e_{\a}), 
\label{mbcs}
\ee
where the relation $M=(N_b-\nu)/2$ was used.  
Substituting the ansatz (\ref{bcs}) 
to the second of the equations (\ref{me}) 
and using the equation (\ref{gapbcs}) 
we obtain the energy:  
\[
E=-\sum_{\a}\e_{\a}+\sum_{\a}C_{\a}\e_{\a}^{2}S(\e_{\a}) 
+\frac{1}{2}\sum_{\a}C_{\a}\e_{\a}S(\e_{\a})(a+b) 
-\frac{(a-b)^2}{4g}. 
\]
Taking into account the equation (\ref{mbcs}) 
after some algebra this expression can 
be represented in the following form: 
\be 
E=-\sum_{\a}\e_{\a}+\sum_{\a}C_{\a}E(\e_{\a}) 
-(N_b+L)\frac{a+b}{2}+
\frac{1}{4g}\left(b-a\right)^2. 
\label{ebcs} 
\ee
Thus the parameters $a$, $b$ determined from the 
equations (\ref{gapbcs}), (\ref{mbcs}) should be 
substituted to the energy (\ref{ebcs}).   
If the parameters $a$, $b$ are fixed, 
if $a\neq0$, the gap in the spectrum of excitations 
will appear and the Bose condensate will be absent.     
   One can further rewrite the equations (\ref{mbcs}), 
(\ref{ebcs}) in order to see the similarity with the 
mean-field (variational) equations presented below.  
   Introducing the notations 
\[
\mu=\frac{|a+b|}{2}, ~~~~~~\d=\frac{|a-b|}{2}, 
\]
the equations (\ref{mbcs}), (\ref{ebcs}) 
for $\mu$, $\d$ at $\nu_{\a}=0$ take the form: 
\[
N_b+L=
\sum_{i}\frac{\e_i+\mu}{\sqrt{(\e_i+\mu)^2-\d^2}},~~~~~~
\sum_{i}\frac{1}{\sqrt{(\e_i+\mu)^2-\d^2}}=\frac{2}{g}, 
\]
\be
E(\mu,\d)=-\sum_i\e_i+
\sum_{i}\sqrt{(\e_i+\mu)^2-\d^2} 
-(N_b+L)\mu +\frac{\d^2}{g}. 
\label{mfbcs}
\ee 
The gap in the excitation spectrum equals $\sqrt{\mu^2-\d^2}$. 
The parameters $\mu$, $\d$ found from the first two 
of the equations (\ref{mfbcs}) should be substituted 
into the energy $E(\mu,\d)$ (\ref{mfbcs}). 
The first two of the equations (\ref{mfbcs}) are equivalent to the 
condition of minimum of the energy $E(a,b)$ 
over the variables $a$, $b$, which in terms of new variables reads 
$\delta E/\delta\mu=0$, $\delta E/\delta\d=0$.  
  Thus the equations (\ref{mfbcs}) are equivalent   
to the equations obtained from the mean-field 
theory (see below). 
The difference of the exact solution with 
the mean-field approach can appear  
only in the weak coupling regime 
in presence of the Bose condensate.

If the minimum of the energy exist, 
the solution of the equations (\ref{mfbcs}) can 
be easily found. 
For example, for the equal-spacing $L$-level model 
with $L\e_1=1$, taking the variations of (\ref{mfbcs}) 
over $\mu$ and $\d$ we get the equations presented in the 
next section. 
The condition of the existence of the solution is 
\[
\sqrt{\mu^2-\d^2}=\frac{1}{2(\rho+2)}
\left(2C-\rho(\rho+2)\right)>0, ~~~~
C=\frac{2+\rho}{e^{2/g}-1} 
\]
Separately the parameters $\mu$, $\d$ can be found from 
the equations 
\[
(\mu^2-\d^2)^{1/2}=(C^2-\d^2)/2C, ~~~~~  
\mu=(C^2+\d^2)/2C.  
\]
For a given density $\rho$ the last equation gives the 
critical value of the coupling constant $g_c$: 
\be 
g_c(\rho)=\frac{2}{\ln(1+2/\rho)}. 
\label{gc}
\ee
For $g>g_c(\rho)$ the solution of the equations  
(\ref{mfbcs}) exist, $|a|>0$, and the gap in the 
energy spectrum $\sqrt{ab}=\sqrt{\mu^2-\d^2}>0$. 
The Bose condensate is absent in this phase. 
For $g=g_c(\rho)$ we have $a=0$ and for 
$g<g_c(\rho)$ the gap closes and the 
solution (\ref{any}) 
with the Bose condensate described above is valid. 
In fact, let us show that the critical value (\ref{gc}) 
coincides with the value of $g$ determined by the 
condition $N_b-N_0<N_b$ in the framework of the 
solution (\ref{any}) by the equation (\ref{bg}). 
One observes that eq.(\ref{m}), (\ref{bg}) 
can be represented in the following form: 
\[
\frac{b}{g}=N_b-\sum_i\left(
\frac{\e_i+b/2}{\sqrt{\e_i(\e_i+b)}}-1\right) 
+\frac{b}{2}\sum_i\frac{1}{\sqrt{\e_i(\e_i+b)}}.     
\]
The cancellation of the first two terms at the right-hand 
side of this equation is equivalent to the condition 
$N'=N_b$ in the framework of the solution (\ref{bcs}), 
while the last sum equals $b/g$ in the framework of the 
solution  (\ref{bcs}) at $a=0$. In fact, from eq.(\ref{mfbcs})  
at $a=0$ ($\mu=\d$) we obtain exactly 
$2/g=\sum_{i}(\e_i(\e_i+b))^{-1/2}$. 
Therefore, two estimates of the critical point 
$g_c$ found from two solutions in the weak and 
the strong coupling limits are coincide.

Let us show that at the value $a=0$ the density 
$R(t)$ given by eq.(\ref{bcs}), which 
was equal to zero at this point, 
is reduced to the density in the weak coupling 
regime (\ref{any}) which is unbounded at $t=0$. 
Due to the term $\sim1/t$ in the parenthesis of eq.(\ref{bcs})   
one can rewrite the density (\ref{bcs}),  
\[
R(t)=\frac{1}{\pi}\sqrt{(t-a)(t-b)}
\left(-\frac{1}{2}\sum_{\a}\frac{S(\e_{\a})}{t-\e_{\a}}\right), 
\]
in the following form: 
\[
R(t)=\frac{1}{\pi}\sqrt{\frac{t-a}{t-b}}
\left(-\frac{1}{g}-\frac{1}{2}\sum_{\a}
\frac{S(\e_{\a})(\e_{\a}-a)}{t-\e_{\a}}\right),  
\]
if the condition $\int d\xi\phi(\xi)=-1/g$ 
(\ref{gapbcs}) is taken into account. 
The last expression coincides with the result 
obtained from the ansatz of the type (\ref{any}) 
if the condition of minimum 
of the energy (\ref{gapbcs}) as a function of $a$, $b$,  
$\sum_{i}(1/E(\e_i))=2/g$ is satisfied. 
However, let us stress that the transition between the 
two phases at the critical point $g_c(\rho)$ is discontinuous.   

    Thus, we have shown that the transition from the 
strong coupling incompressible phase with the gap to the phase 
with the Bose condensate and the Bogoliubov- type 
spectrum of phonons takes place at the coupling 
constant $g=g_c(\rho)$ (\ref{gc}). 
At this point the condensate is equal to zero, 
$N_0=0$, but at $g<g_c(\rho)$ the condensate 
increases according to the equation (\ref{b}) 
until the value $N_0=N_b$ is reached at $g=0$. 
Let us note that qualitatively these results 
are valid for the model with an arbitrary 
degeneracy of energy levels $\Omega_{\a}$ 
and an arbitrary level spacing. 
Numerically the dependence $g_c(\rho)$ will 
have the different form.    
The limiting case of the solution at $|g|\to\infty$ coincides 
with the solution of the one-level model in this limit.

\vspace{0.2in}

  {\bf 4. Mean-field solution.}  

\vspace{0.1in} 

Here we consider the mean field or variational 
solution of the pairing model (\ref{H}) for the case of 
attraction:  
\be
H=\sum_{i=0}^{L-1}\e_i n_i -gB^{+}B^{-}, ~~~~g>0. 
\label{Hat}
\ee
Let us describe the mean-field approach for the model 
(\ref{Hat}) and find the range of the parameters for which the 
solution is exact in the thermodynamic limit. 
The mean-field Hamiltonian has the form  
\be
H_{MF}=\sum_{i}(\e_i+\mu)n_i+\d\sum_{i}(b_i^{+}+b_i)-\mu N_b
+\frac{\d^2}{g}, 
\label{mfH}
\ee
where $\mu$ is the chemical potential and the variational 
parameter $\d$ is real. 
   The expression (\ref{mfH}) can be considered in a sense of the 
Hubbard - Stratanovich transformation in the functional integral 
which can also be used to establish the validity of the mean-field 
theory.  For the Hamiltonian (\ref{Hat}) 
the mean field theory (\ref{mfH}) is equivalent to the variational 
procedure with the trial wave function analogous to the BCS 
wave function.  
     It is well known that the variational solution for the BCS 
Hamiltonian is exact in the thermodynamic limit (for example, 
see \cite{Bardeen}). 
In contrast to the BCS case, for the 
bosonic model one has to introduce the chemical potential 
in order to fix the particle number.  
   We show that in some range of parameters, 
at $g>g_c(\rho)$, the variational solution for the bosonic 
pairing model coincides with the exact solution presented above. 
 At $g<g_c(\rho)$ the naive mean field solution is 
not correct. However, for our model one can modify the mean field 
(variational) approach taking into account the 
Bose condensation to obtain the exact results 
presented above in the whole range of parameters 
(except the extremely small coupling constant $g\rho\sim\e_1$).

Each of the quadratic Hamiltonians $H_i$ in the sum (\ref{mfH}) 
can be diagonalized by means of the Bogoliubov transformation. 
For each site $i$ introduce the new Bose creation and annihilation 
operators $\chi_{1,2}$, $\chi_{1,2}^{+}$ according to 
\[
\phi_1^{+}=c\chi_1^{+}+s\chi_2,~~~~\phi_2^{+}=c\chi_2^{+}+s\chi_1, 
\]
where the coefficients $c_i$, $s_i$ are assumed to be real, 
\[
c_i^2-s_i^2=1,~~~c_i=\ch(\phi_i), ~~s_i=\sh(\phi_i). 
\]
The expectation values of the particle number $n_i$ and 
the energy $H_i$ in the ground state are 
\[
\la n_i\ra=2s_i^2, ~~~~
\la H_i\ra=(\e_i+\mu)(c_i^2+s_i^2-1)+\d2c_{i}s_i.  
\]
The condition of cancellation of the terms 
$\chi_1\chi_2$ and $\chi_1^{+}\chi_2^{+}$ takes the form: 
\[
\frac{2c_{i}s_{i}}{c_i^2+s_i^2}=\th(2\phi_i)=-\frac{\d}{\e_i+\mu}.  
\]
Thus we obtain the expressions for the energy and the 
number of particles as a functions of the parameter $\d$ and the 
chemical potential $\mu$: 
\[
 E_{MF}(\d)=\sum_{i}\left(\sqrt{(\e_i+\mu)^2-\d^2}(1+n^{\chi}_i)
-(\e_i+\mu)\right)-\mu N_b+\frac{\d^2}{g},~~~
\]
\be
 N_b=\sum_{i}\left(\frac{|\e_i+\mu|}{\sqrt{(\e_i+\mu)^2-\d^2}}
(1+n^{\chi}_i)-1\right), 
\label{mf}
\ee
where the operator $n^{\chi}_i$ equals 
\[
n^{\chi}_i=\chi_{1i}^{+}\chi_{1i}+\chi_{2i}^{+}\chi_{2i}, ~~~~
\nu_i\sigma_i=\chi_{1i}^{+}\chi_{1i}-\chi_{2i}^{+}\chi_{2i}. 
\]
The parameters $\d$ and $\mu$ should be determined from the 
condition of minimum of $E_{MF}(\d)$ (\ref{mf}) with the condition 
of fixed number of particles $N_b$. 
From eq.(\ref{mf}) the excitation energy 
$E_i=\sqrt{(\e_i+\mu)^2-\d^2}$. 
The ground state corresponds to the quantum numbers $n^{\chi}_i=0$, 
or, equivalently, to the state $|0\ra_{\chi}$ annihilated by the 
operators $\chi_{1,2}$: 
\[
 \chi_{1i}|0\ra_{\chi}=0, ~~~\chi_{2i}|0\ra_{\chi}=0,~~~ 
i=0,\ldots L-1. 
\]
In terms of the initial operators $\phi_1^{+}$, $\phi_2^{+}$ this 
state can be represented as 
\be
|0\ra_{\chi}=
\prod_{i}e^{\alpha_{i}(\phi_{1i}^{+}\phi_{2i}^{+})}|0\ra, ~~~
\alpha_i=\frac{s_i}{c_i}=\th(\phi_i), 
\label{a}
\ee
where $|0\ra$ is the vacuum with respect to the initial operators: 
$\phi_{1i}|0\ra=0$, $\phi_{2i}|0\ra=0$. In fact, for each $i$ 
for the state $|\a\ra=\exp(\a\phi_1^{+}\phi_2^{+})\vac$ one finds 
\[
(\phi_1-\a\phi_2^{+})|\a\ra=0, 
~~~~|\a\ra=e^{\a(\phi_1^{+}\phi_2^{+})}\vac.  
\]
Substituting the expressions for the operators $\phi_1$, $\phi_2$, 
one can see that the state (\ref{a}) is annihilated by
the operators $\chi_1$, $\chi_2$ provided the condition 
$\a=s/c=\th(\phi)$  
is satisfied. 
     The excited 
states can be constructed from the state (\ref{a}) 
by action of the operators $\chi_1^{+}$, $\chi_2^{+}$: 
\[
(\chi_{1}^{+})^{n_{1}}(\chi_{2}^{+})^{n_{2}}|0\ra_{\chi}, 
 ~~~~\nu=|n_{1}-n_{2}|. 
\]
Let us show that for the model (\ref{Hat}) 
the mean-field theory approach is equivalent 
to the variational procedure with the trial variational wave 
function of the form (\ref{a}). Although this wave function 
does not correspond to a definite particle number, it can be 
fixed in an average as in the usual BCS theory, 
which is justified in the continuum limit.   
Expectation value of the Hamiltonian (\ref{Hat}) 
over the state (\ref{a}) as a function of the variational 
parameters $\phi_i$ takes the form: 
\be
E=\sum_{i}(\e_i+\mu)(2s_i^2)+
g\left(\sum_{i}c_{i}s_i\right)^2 -\mu N_b.  
\label{evar}
\ee
Taking the variation of (\ref{evar}) with respect to 
$\phi_i$ one finds the equations presented above with  
\[
\d=g\sum_{i}(c_{i}s_i).
\]
The average particle number $N_b=\sum_{i}2s_i^2$. 
The existence of the condensate means 
$\phi_0\to\infty$. Substituting this value to the 
right -hand side of eq.(\ref{evar}) and 
assuming $N_0=N_b$, one finds $\mu=g\rho/2$,  
and the spectrum $E(\e)=\sqrt{\e(\e+g\rho)}$ in agreement 
with the Bogoliubov approximation. 
Thus although the Bogoliubov approximation 
corresponds to the variational estimate of the energy, 
in general, it does not correspond to the minimum of 
the energy on the class of the wave functions (\ref{a}). 
However we show that in the weak coupling limit the 
naive variational approach does not lead to the correct results 
while the Bogoliubov approximation gives the exact results 
in the weak coupling limit if the density is not too small 
($1\ll\rho\ll 1/g$).

In the strong coupling regime $g>g_c(\rho)$ 
the equations (\ref{mf}) coincide with the 
exact equations obtained in section 6. 
The equation $\delta E_{MF}(\d)/\delta\d=0$ 
together with the second of the equations 
(\ref{mf}) allows one to find the parameters 
$\mu$, $\d$, the occupation numbers and the gap 
in the energy spectrum. In particular, for the 
equal-spacing $L$ level model with $L\e_1=1$ 
the equations take the form: 
\be
\ln\left(1+\mu+\sqrt{(1+\mu)^2-\d^2}\right)- 
\ln\left(\mu+\sqrt{\mu^2-\d^2}\right)=\frac{2}{g},  
\label{min} 
\ee
\be
\rho+1=\sqrt{(\mu+1)^2-\d^2}-\sqrt{\mu^2-\d^2},  
\label{number}
\ee
which have the solution found in section 6. 
The results are in agreement with the exact solution.

Let us see if the exact solution in the 
weak coupling case $g<g_c$ can be obtained in 
the framework of the mean-field (variational) approach. 
To get the expectation value $\la n_0\ra=N_0$ of order 
$N_b$, one should take $\d=\mu+\delta$, where 
$\delta$ is the small parameter of order $1/L$.  
Substituting the value $\mu=\d$ into one of the  
equations (\ref{min}), (\ref{number}) one finds 
that the result for $\d$ contradicts the exact solution. 
Thus the naive mean field approach fails for the 
region of the parameters where the solution of the 
equations $\delta E_{MF}/\delta\mu=0$, 
$\delta E_{MF}/\delta\d=0$ does not exist.    
To get the correct results, one should 
use the following method.  
First, substitute the parameter $\mu=\d$ 
into $E_{MF}(\mu,\d)\to E_{MF}(\d,\d)$ (\ref{mf}).   
Then the solution of the equation 
$\delta E_{MF}/\delta\d=0$ gives the results 
in agreement with exact solution of section 6. 
In fact, one can see that the variation of this 
function leads to the equation (\ref{m}), 
which in the framework of the exact solution 
was used to determine the parameter $b$. 
   The validity of this method can be shown 
in the same way as for the usual BCS model.  
In the framework of the functional integral approach    
the factor $L$ (the volume) appears in the exponent 
in front of the action if the condensate is absent.  
 To take into account the condensate one can 
introduce the $\delta$- function of the form 
\[
\delta(n_0(\mu,\d)-N_b+N'(\d)), 
\]
where the function $n_0(\mu,\d)=\la n_0\ra$ is 
given by eq.(\ref{mf}) for $i=0$ and the function 
$N'(\d)$ is determined by the sum 
$\sum_{i\neq0}\la n_i\ra$, with $\la n_i\ra$ 
given by eq.(\ref{mf}) with $\mu=\d$.  
This factor will give $\mu=\d$ with the accuracy of 
order $1/N_b$ and remove the integration over $\mu$. 
If the saddle point for the remaining integration 
over $\d$ exist and gives the value 
$N'(\d)<N_b$, which indeed takes place, 
the solution is exact in the thermodynamic limit. 
The particle number is correctly fixed 
within this approach.    
The same can also be shown using the trial variational wave 
function of the form $|N_0,\phi_1,\ldots\phi_{L-1}\ra$, 
where $N_0$ and $\phi_i$, $i\neq0$ are the variational 
parameters. 
Thus the modified mean-field approach is valid 
in the whole range of the parameters with the 
exception of the extremely small coupling constant 
$g\rho\sim\e_1$, when the value of $N'$, the number 
of particles out of the condensate, becomes of 
order of unity.

\vspace{0.2in}

     {\bf Conclusion.}

\vspace{0.1in} 

 In the present paper we have shown that the discrete-state BCS-type 
pairing models for bosons can be considered as a quasiclassical limit 
of the eigenvalue problem of the general transfer matrix in the 
framework of the algebraic Bethe ansatz method. 
We introduced the new pairing model for bosons corresponding to the 
attractive pairing interaction. It was shown that the weak coupling 
phase, $g<g_c$, is characterized by the Bose condensation and the 
Bogoliubov-type spectrum of phonons. In the strong coupling phase 
at $g>g_c$ the Bose condensate is absent and there is a gap 
in the excitation spectrum.   
Note that the transition of this type from the incompressible 
Mott insulating phase to the superfluid phase is 
usually expected in the Bose Hubbard model. 
We have shown that naive variational 
approach is not applicable in the weak coupling limit 
at $g<g_c$, when the condensate fraction exist.  
 However, for our model one can modify the 
variational procedure taking into account the 
condensate fraction to obtain the exact solution in the  
whole range of parameters. 
   The Bogoliubov approximation gives the 
correct results in agreement with the exact solution 
in the limit $g\ll1$ and $1\ll\rho\ll 1/g$, such that 
the parameter $g\rho\ll1$, i.e. when the parameter 
$b=g\rho$ (see eq.(\ref{bg})).     
   The proposed model with an attractive pairing interaction 
can be interesting both in the context of applications to the 
finite systems of the confined bosons and for studying the 
phenomenon of superfluidity in the exactly- solvable model.

\vspace{0.2in} 

     {\bf Acknowledgments.}

The author is grateful to V.A.Rubakov for useful remarks. 
This work was supported in part by RFFI Grant 
NSh-2184.2003.2.

\end{document}